\documentclass[12pt,a4paper]{extarticle}
\usepackage[utf8]{inputenc}
\usepackage[english]{babel}
\usepackage[OT1]{fontenc}
\usepackage{amsmath}
\usepackage{mathrsfs}
\usepackage{amsfonts}
\usepackage{hyperref}
\usepackage{amssymb}
\usepackage{eufrak}
\usepackage{graphicx}
\usepackage[left=2.2cm,right=2.2cm,top=2cm,bottom=2cm]{geometry}
\allowdisplaybreaks
\begin{document}

\begin{flushright}

\end{flushright}

\vspace{0.5cm}
\begin{center}
{\large\bf Covariant structure constants for deformed oscillator algebra}

\vspace{1 cm}

{\bf A.V.Korybut}\\
\vspace{0.5 cm} {\it
Moscow Institute of Physics and Technology (State University),\\
Institutskii per. 9, 141700, Dolgoprudny, Moscow region, Russia }

\vspace{0.6 cm}
akoribut@gmail.com\\
\end{center}

\vspace{0.4 cm}

\begin{abstract}
\noindent $sl_2$-covariant expressions for structure constants of the associative algebra of deformed oscillators $Aq\left(2,\nu\right)$ are obtained.
\end{abstract}
\newpage
\tableofcontents
\newpage
\section{Introduction}

Deformed oscillator algebra with the generators $y_\alpha$ ($\alpha =1,2$) and 
$\mathcal{K}$ obeying relations
\begin{equation}
\left[y_\alpha,y_\beta\right]=2i\epsilon_{\alpha \beta} \left(1+\nu \mathcal{K}\right), \;\; \lbrace y_\alpha, \mathcal{K} \rbrace=0, \;\; \mathcal{K}^2=1,
\end{equation}
where $\nu\in \mathbb{C}$ is an arbitrary parameter, was originally considered by Wigner in its particular representation\cite{Wigner}. In \cite{sphere} and \cite{jetp}. it was argued that it can be interpreted as a higher-spin (HS) algebra in $2+1$ dimensions. In a slightly different realization thesame  HS algebra was considered in \cite{Stelle},\cite{Fradkin1}. More generally  deformed oscillator algebra determines the form of non-linear equations in any dimension \cite{nhs},\cite{nhs2}. Also this algebra appears in context of $2d$ conformal field theory \cite{Pope},\cite{VasBerg1},\cite{VasBerg2}. Interest to this algebra increased recently in the context of $AdS_3/CFT_2$ correspondence \cite{GabMin}. Analysis of these problems is complicated in particular by the absence of convenient expression for the product in this algebra.

 Structure constants for bosonic HS Lie algebra were found by Pope, Shen, Romans in \cite{Pope} and for  HS Lie superalgebra  by Fradkin and Linetsky in \cite{Fradkin2}. In \cite{Pope} the Lone-Star product was introduced as the associative product underlying $hs\left(\lambda\right)$. All the results concerning structure constants in these papers were given in conformal basis
\begin{equation}
V^s_n=\left(-1\right)^{s-1-n} \dfrac{\left(n+s-1\right)!}{\left(2s-2\right)!}\left[J_-,\dots \left[J_-,\left[J_-,J_+^{s-1}\right] \right] \right],
\end{equation}
where $J_-,J_+$ are generators of $sl_2$ with commutation rules
\begin{equation}
\left[J_+,J_-\right]=2 J_0, \;\;\;\;\;\; \left[J_\pm,J_0\right]=\pm J_\pm.
\end{equation}
Recently the same problem was analyzed in \cite{Young}. For functions
\begin{multline}
\displaystyle
L\left(\xi\right)=\sum_{n=0}^\infty \dfrac{1}{n!} \xi^{\alpha_1 \tilde{\alpha}_1} \dots \xi^{\alpha_n \tilde{\alpha}_n} y_{(\alpha_1}y_{\tilde{\alpha}_1}\dots y_{\alpha_n}y_{\tilde{\alpha}_n )}, \\
L\left(\eta\right)=\sum_{n=0}^\infty \dfrac{1}{n!} \eta^{\alpha_1 \tilde{\alpha}_1} \dots \eta^{\alpha_n \tilde{\alpha}_n} y_{(\alpha_1}y_{\tilde{\alpha}_1}\dots y_{\alpha_n}y_{\tilde{\alpha}_n )},
\end{multline}
where $\xi^{\alpha_1 \tilde{\alpha}_1}$ and $\eta^{\alpha_1 \tilde{\alpha}_1}$ are symmetric tensors, the product is
\begin{equation}
\displaystyle
L\left(\xi\right) \ast L\left(\eta
\right)=\sum_{n=0}^{\infty} {}_2F_1\left(\begin{matrix} n+\frac{3-\nu}{2} & n+\frac{1+\nu}{2} \\
 & n+\frac{3}{2} \end{matrix} \;\;  ; \;\; -\dfrac{1}{4} \phi \right) 
 \dfrac{1}{n!} \zeta^{\alpha_1 \tilde{\alpha}_1} \dots \zeta^{\alpha_n \tilde{\alpha}_n} y_{(\alpha_1}y_{\tilde{\alpha}_1}\dots y_{\alpha_n}y_{\tilde{\alpha}_n )},
\end{equation}
where $\zeta^{\alpha \beta}=\xi^{\alpha \beta}+\eta^{\alpha \beta}+\xi^{\alpha \mu} \eta^{\nu \beta} \epsilon_{\mu\nu}$ and $\phi=\xi^{\alpha_1 \alpha_2} \eta^{\beta_1 \beta_2} \epsilon_{\alpha_1 \beta_1} \epsilon_{\alpha_2 \beta_2}$.

HS algebras in higher dimensions admit star product operation
\begin{equation}
\displaystyle \left(f\ast g\right)\left(y\right)=\dfrac{1}{\left(2 \pi\right)^{2p}} \int d^{2p}u \; d^{2p}v \exp\left(iu_{\mu} v^{\mu}\right) f\left(y+u\right) g\left(y+v\right) \; .
\end{equation}
Benefit of this operation is that in r.h.s. there is the integral of functions of commuting variables. No counterpart is known for deformed oscillator case.  But a step forward was made, in star product initial and final functions are power series of oscillators with totally symmetric tensors (Weyl ordering)

\begin{equation}
\displaystyle
f\left(y\right)=\sum_{n=0}^{\infty} f^{\alpha\left(n\right)}\underbrace{y_{\alpha}\dots y_\alpha}_n, \;\;\;\;\;\;\;\; f^{\alpha\left(n\right)}=f^{\left(\alpha_1 \dots \alpha_n\right)}.
\end{equation}
$f^{\alpha\left(n\right)}$ - totally symmetric tensor of rank n.

\begin{equation}
\label{prod}
\displaystyle
f^{\alpha\left(n\right)} \underbrace{y_\alpha \dots y_\alpha}_n \ast g^{\beta\left(m\right)} \underbrace{y_\beta \dots y_\beta}_m = f^{\alpha\left(n\right)} g^{\beta\left(m\right)} \sum_{p=0}^{min\left( m,n \right)} \tilde{A}\left(m,n,p,\nu\right) \left(\epsilon_{\alpha \beta} \right)^p \underbrace{\overbrace{y_\alpha \dots y_\alpha}^{m-p} \overbrace{y_\beta \dots y_\beta}^{n-p}}_{m+n-2p}
\end{equation}
Oscillators are totally symmetrized. Expression for $\tilde{A}$ for different cases is given in the next section.

The paper is organized as follows: in section 2 structure constants and covariant expression for product in $Aq\left(2,\nu\right)$ are given, in section 3 we derive associativity condition, checked in appendix, in sections 4 and 5 structure constants for other cases are derived, in section 5 conclusion.

\section{Structure constants for $Aq\left(2,\nu\right)$}

To clarify issue with Klein operator which obviously should appear in r.h.s. of Eq. (\ref{prod}) we build a pair of projectors:

\begin{equation}
\Pi_{\pm}=\dfrac{1 \pm \mathcal{K}}{2}.
\end{equation}
We can get rid of Klein operator in order to make expressions shorter. For this purpose we redefine the product

\begin{equation}
f\left(y\right) *g \left(y \right) :=f\left(y\right) *g \left(y \right) \Pi_{\pm}.
\end{equation}

\noindent In all the computations below multiplications with the projector is assumed. We will use $\Pi_+$, to get the answer for $\Pi_-$ one will need simply to change the sign before $\nu$. General expression for $\tilde{A}\left(m,n,p,\nu\right)$ is a bit involved and it is convenient to separate different cases. The final result for structure constants for $Aq\left(2,\nu\right)$ is

\textit{Even $\times$ Even}

\begin{equation}
\label{Aprod}
A \left(m,n,p,\nu\right)=i^p \dfrac{m! n!}{\left(m-p\right)!\left(n-p\right)! p!} {}_4 F_3\left[\begin{matrix} 1-\frac{\nu}{2}& \frac{\nu}{2} &\frac{-p}{2}& \frac{1-p}{2}\\
\frac{1-m}{2}& \frac{1-n}{2} &\frac{m+n-2p+3}{2}\end{matrix};1\right],
\end{equation}

\textit{Odd $\times$ Odd}

\begin{multline}
B\left(m+1,n+1,p,\nu\right)=A\left(m,n,p,-\nu\right)+i \left(m+n-2p+3+\nu\right)A\left(m,n,p-1,-\nu\right)+\\
+i^2 \left(m-p+2\right)\left(n-p+2\right) \dfrac{m+n-2p+5+\nu}{m+n-2p+5} \dfrac{m+n-2p+3+\nu}{m+n-2p+3} A\left(m,n,p-2,\nu\right) ,
\end{multline}

\textit{Even $\times$ Odd}

\begin{equation}
C\left(m,n+1,p,\nu\right)=A\left(m,n,p,-\nu\right)+i\left(m-p+1\right)\dfrac{m+n-2p+3+\nu}{m+n-2p+3} A\left(m,n,p-1,-\nu\right),
\end{equation}

\textit{Odd $\times$ Even}

\begin{equation}
D\left(m+1,n,p,\nu\right)=A\left(m,n,p,\nu\right)+i\left(n-p+1\right)\dfrac{m+n-2p+3-\nu}{m+n-2p+3} A\left(m,n,p-1,\nu\right),
\end{equation}
where $m,n$ - are even numbers. For general

\begin{equation}
\displaystyle
f\left(y\right)=\sum_{m=0}^{\infty} f^{\alpha\left(m\right)}\underbrace{y_{\alpha}\dots y_\alpha}_m, \;\;\;\;\;
g\left(y\right)=\sum_{n=0}^{\infty} g^{\beta\left(n\right)}\underbrace{y_{\beta}\dots y_\beta}_n
\end{equation}
the product is
\begin{multline}
\displaystyle
f\left(y\right) \ast g\left(y\right)= \sum_{k=0}^\infty \sum_{p=0}^\infty \left( f^{\alpha\left(2k\right)}g^{\beta\left(2p\right)} \sum_{l=0}^{min\left(2k,2p\right)} A\left(2k,2p,l,\nu\right)\left(\epsilon_{\alpha \beta} \right)^l \underbrace{\overbrace{y_\alpha \dots y_\alpha}^{2k-l} \overbrace{y_\beta \dots y_\beta}^{2p-l}}_{2k+2p-2l}+ \right. \\ \left.
f^{\alpha\left(2k+1\right)}g^{\beta\left(2p+1\right)} \sum_{l=0}^{min\left(2k+1,2p+1\right)} B\left(2k+1,2p+1,l,\nu\right)\left(\epsilon_{\alpha \beta} \right)^l \underbrace{\overbrace{y_\alpha \dots y_\alpha}^{2k+1-l} \overbrace{y_\beta \dots y_\beta}^{2p+1-l}}_{2k+2p+2-2l}+ \right. \\ \left.
f^{\alpha\left(2k\right)}g^{\beta\left(2p+1\right)} \sum_{l=0}^{min\left(2k,2p+1\right)} C\left(2k,2p+1,l,\nu\right)\left(\epsilon_{\alpha \beta} \right)^l \underbrace{\overbrace{y_\alpha \dots y_\alpha}^{2k-l} \overbrace{y_\beta \dots y_\beta}^{2p+1-l}}_{2k+2p+1-2l}+\right. \\ \left.
f^{\alpha\left(2k+1\right)}g^{\beta\left(2p\right)} \sum_{l=0}^{min\left(2k+1,2p\right)} D\left(2k+1,2p,l,\nu\right)\left(\epsilon_{\alpha \beta} \right)^l \underbrace{\overbrace{y_\alpha \dots y_\alpha}^{2k+1-l} \overbrace{y_\beta \dots y_\beta}^{2p-l}}_{2k+2p+1-2l}
\right).
\end{multline}

\section{The associativity condition}
In this section the associativity condition for bosonic structure constants (\textit{Even $\times$ Even}) is derived. Actually it is enough to consider only this case because of the way other structure constants were derived. Product of two monomials can be written in the form:

\begin{equation}
\label{m times n}
f^{\alpha \left( m \right)} \overbrace{y_\alpha \dots y_\alpha}^m \ast g^{\beta \left( n \right)}
\overbrace{y_\beta \dots y_\beta}^n = f^{\alpha \left( m \right)}g^{\beta \left( n \right)}
\sum^{min \left(m ,n \right)}_{p=0} A\left( m,n,p, \nu \right) \left(\epsilon_{\alpha \beta} \right)^{p} \underbrace{\overbrace{y_\alpha \dots y_\alpha}^{m-p} \overbrace{y_\beta \dots y_\beta}^{n-p}}_{m+n-2p},
\end{equation}
where $A\left( m,n,p, \nu \right)$ are structure constants for \textit{Even $\times$ Even} case.  In this section we consider only even monomials. We can rewrite previous equation in a slightly different form. Using associativity

\begin{multline}
\label{m times n-2}
\displaystyle f^{\alpha \left( m \right)} \overbrace{y_\alpha \dots y_\alpha}^m \ast g^{\beta \left( n-2 \right) \gamma \gamma}
\overbrace{y_\beta \dots y_\beta}^{n-2} y_\gamma y_\gamma
=\\
= f^{\alpha \left( m \right)}g^{\beta \left( n-2 \right) \gamma \gamma}
\left( \sum^{min \left(m ,n-2 \right)}_{p=0} A\left( m,n-2,p, \nu \right) \left(\epsilon_{\alpha \beta} \right)^{p} \underbrace{\overbrace{y_\alpha \dots y_\alpha}^{m-p} \overbrace{y_\beta \dots y_\beta}^{n-2-p}}_{m+n-2-2p} \right) y_\gamma y_\gamma.
\end{multline}

\noindent where associativity of product is used. We can multiply $ m \times n$ directly or $m \times \left(n-2\right) \times 2$. Due to associativity result should be the same. To express it in the form of equation we have to symmetrize oscillators with indices $\gamma$. Other oscillators are already symmetrized.

\begin{multline}
\displaystyle
\underbrace{\overbrace{y_\alpha \dots y_\alpha}^{m-p} \overbrace{y_\beta \dots y_\beta}^{n-2-p}}_{m+n-2-2p} y_\gamma y_\gamma= \\
= \left[ \overbrace{y_{( \alpha} \dots y_\alpha}^{m-p} \overbrace{y_\beta \dots y_\beta}^{n-2-p} y_{\gamma )} + \dfrac{2i}{2} \dfrac{\left(m+n-2p-2\right)}{m+n-1-2p} \left(m+n-2p-1-\nu\right) \xi \; \epsilon_{\alpha\gamma} \overbrace{y_{(\alpha} \dots y_\alpha}^{m-1-p}\overbrace{y_{\beta} \dots y_{\beta)}}^{n-2-p} \right]y_\gamma
\end{multline}
The factor
\begin{equation}
\dfrac{2i}{2} \dfrac{\left(m+n-2p-2\right)}{m+n-1-2p} \xi ,
\end{equation}
which appears in the second term containing $\epsilon_{\alpha \gamma}$ before the contraction consists of two parts.
The first part
\begin{equation}
\dfrac{2i}{2} \dfrac{\left(m+n-2p-2\right)}{m+n-1-2p} \left(m+n-2p-1-\nu\right).
\end{equation}
Results from symmetrization of $m+n-2p-2$ totally symmetric oscillators with one more oscillator.
\begin{equation}
\displaystyle
\underbrace{y_\rho \dots y_\rho}_{m+n-2-2p} y_\gamma y_\gamma=\left[\underbrace{y_{(\rho} \dots y_\rho}_{m+n-2-2p} y_{\gamma)}+\dfrac{2i}{2} \dfrac{\left(m+n-2p-2\right)}{m+n-1-2p} \left(m+n-2p-1-\nu\right)\underbrace{y_\rho \dots y_\rho}_{m+n-3-2p} \epsilon_{\rho \gamma}\right] y_\gamma.
\end{equation}
The second part, namely $\xi$ is purely combinatorial. It appears because $y_\beta$ and $y_\gamma$ are already symmetrized. To derive the value of $\xi$ we should understand symmetries of the initial expression

\begin{equation}
f^{\alpha_1 \dots \alpha_{m-p} \mu_1 \dots \mu_p}g^{\tilde{\mu}_1 \dots \tilde{\mu}_p \beta_1 \dots \beta_{n-2-p} \gamma \gamma} \epsilon_{\mu_1 \tilde{\mu}_1} \dots \epsilon_{\mu_p \tilde{\mu}_p} \overbrace{y_{( \alpha_1} \dots y_{\alpha_{m-p}}}^{m-p} \overbrace{y_{( \beta_1} \dots y_{\beta_{n-2-p})}}^{n-2-p} y_\gamma y_\gamma
\end{equation}
and contraction

\begin{equation}
f^{\alpha_1 \dots \alpha_{m-p} \mu_1 \dots \mu_p}g^{\tilde{\mu}_1 \dots \tilde{\mu}_p \beta_1 \dots \beta_{n-2-p} \gamma \gamma} \epsilon_{\mu_1 \tilde{\mu}_1} \dots \epsilon_{\mu_p \tilde{\mu}_p} \overbrace{y_{( \alpha_1} \dots y_{\alpha_{m-p}}}^{m-p} \overbrace{y_{( \beta_1} \dots y_{\beta_{n-2-p})}}^{n-2-p} \; \epsilon_{\beta_1 \gamma} \; y_\gamma .
\end{equation}
The number of permutation which give zeros is

\begin{equation}
\left(n-2-p\right) \dfrac{\left(m+n-2p-3\right)!}{\left(m-p\right)!}.
\end{equation}
The number of permutation that give non-zero contractions is

\begin{equation}
\xi=\dfrac{\left(m-p\right)!}{\left(m+n-2p-2\right)!}\left[\dfrac{\left(m+n-2p-2\right)!}{\left(m-p\right)!}-\left(n-2-p\right) \dfrac{\left(m+n-2p-3\right)!}{\left(m-p\right)!} \right]=\dfrac{m-p}{m+n-2p-2}.
\end{equation}
We have to repeat the same operations with one remaining oscillator $y_\gamma$. This gives Eq. ~\ref{m times n-2}

\begin{multline}
\displaystyle
f^{\alpha (m )} g^{\beta (n)} \sum^{min \left(m ,n-2 \right)}_{p=0} A\left( m,n-2,p, \nu \right) \left(\epsilon_{\alpha \beta}\right)^p \left[ \overbrace{y_{(\alpha} \dots y_\alpha}^{m-p} \overbrace{y_\beta \dots y_{\beta)}}^{n-p}+ \right. \\ \left. + \left(2 i \right) \left(m-p\right) \epsilon_{\alpha \beta} \overbrace{y_{(\alpha} \dots y_\alpha}^{m-p-1} \overbrace{y_\beta \dots y_{\beta)}}^{n-p-1} +\right.\\
\left.\dfrac{\left(2 i\right)^2}{4} \left(\epsilon_{\alpha \beta}\right)^2\left(m-p\right)\left(m-p-1\right)\dfrac{m+n-2p-1-\nu}{m+n-2p-1}\dfrac{m+n-2p-3+\nu}{m+n-2p-3} \overbrace{y_{(\alpha} \dots y_\alpha}^{m-p-2} \overbrace{y_\beta \dots y_{\beta)}}^{n-p-2} \right].
\end{multline}

\begin{equation}
\label{eq0}
A \left(m,n,0,\nu\right)=A \left(m,n-2,0,\nu\right),
\end{equation}

\begin{equation}
\label{eq1}
A \left(m,n,1,\nu\right)=A \left(m,n-2,1,\nu\right)+2i m A \left(m,n-2,0,\nu\right),
\end{equation}

\begin{multline}
A \left(m,n,p,\nu\right)=A \left(m,n-2,p,\nu\right)+2i \left(m-p+1\right) A \left(m,n-2,p-1,\nu\right)+\\
+i^2 A \left(m,n-2,p-2,\nu\right)\left(m-p+2\right)\left(m-p+1\right) \dfrac{m+n-2p+3-\nu}{m+n-2p+3}\; \dfrac{m+n-2p+1+\nu}{m+n-2p+1} \, ,
\label{main}
\end{multline}
where $p=2, \dots , n-2$,
\begin{multline}
A \left(m,n,n-1,\nu\right)=2i \left(m-n+2\right) A \left(m,n-2,n-2,\nu\right)+\\
+i^2 A \left(m,n-2,n-3,\nu\right)\left(m-n+3\right)\left(m-n+2\right) \dfrac{m-n+5-\nu}{m-n+5}\; \dfrac{m-n+3+\nu}{m-n+3},
\label{eq2}
\end{multline}

\begin{multline}
A \left(m,n,n,\nu\right)=
i^2 A \left(m,n-2,n-2,\nu\right)\left(m-n+2\right)\left(m-n+1\right) \dfrac{m-n+3-\nu}{m-n+3}\; \dfrac{m-n+1+\nu}{m-n+1}.
\label{eq3}
\end{multline}
Together with structure constant $A\left(m,2,p,\nu\right)$  these equations completely define structure constants, as they determines $A\left(m,n,p,\nu\right)$ in terms of $A\left(m,2,p,\nu\right)$.

\section{$Odd \times Odd$}
To derive structure constants for odd monomials it is suffices to use associativity and constants for the bosonic case.

\begin{multline}
\label{Ferm}
f\ast g=
f^{\alpha \left( m+1 \right)} \overbrace{y_\alpha \dots y_\alpha}^{m+1} \ast g^{\beta \left( n+1 \right)}
\overbrace{y_\beta \dots y_\beta}^{n+1} = \\
= f^{\alpha \left( m+1 \right)}g^{\beta \left( n+1 \right)}
\sum^{min \left(m ,n \right)+1}_{p=0} B\left( m+1,n+1,p, \nu \right) \left(\epsilon_{\alpha \beta} \right)^{p} \underbrace{\overbrace{y_\alpha \dots y_\alpha}^{m+1-p} \overbrace{y_\beta \dots y_\beta}^{n+1-p}}_{m+n-2p+2},
\end{multline}

\noindent where $m,n$ are even numbers.

\begin{multline}
f^{\alpha \left( m+1 \right)} \overbrace{y_\alpha \dots y_\alpha}^{\left(m+1\right)} \ast g^{\beta \left( n+1 \right)}
\overbrace{y_\beta \dots y_\beta}^{\left(n+1\right)} =
f^{\alpha \alpha \left( m\right)} g^{\beta \left( n \right) \beta} y_{\alpha} \ast \overbrace{y_\alpha \dots y_\alpha}^{\left(m\right)} \ast \overbrace{y_\beta \dots y_\beta}^{\left(n\right)} \ast y_{\beta}=\\
=f^{\alpha \alpha \left( m\right)} g^{\beta \left( n \right) \beta} y_\alpha \; \ast \; 
\sum^{min \left(m ,n \right)}_{p=0} A\left( m,n,p, -\nu \right) \left(\epsilon_{\alpha \beta} \right)^{p} \underbrace{\overbrace{y_\alpha \dots y_\alpha}^{m-p} \overbrace{y_\beta \dots y_\beta}^{n-p}}_{m+n-2p} \; \ast \; y_\beta.
\end{multline}
Now one has to perform symmetrization inserting proper combinatorial factors. Because the procedure is pretty much the same as in deriving Eq.~(\ref{main}) we give only the final answer

\begin{multline}
\displaystyle
 f \ast g= f^{\alpha\left(m+1\right)}g^{\beta\left(n+1\right)} \sum_{p=0}^{min\left(m,n\right)} A\left(m,n,p,-\nu\right) \left( \left(\epsilon_{\alpha \beta}\right)^p  y_{\alpha\left(m+1-p\right)} y_{\beta\left(n+1-p\right)} + \right. \\ \left.
+ i\left(\epsilon_{\alpha \beta}\right)^{p+1} \left(m+n-2p+1+\nu\right) y_{\alpha\left(m-p\right)} y_{\beta\left(n-p\right)} + \right. \\ \left.
+i^2 \left(\epsilon_{\alpha \beta}\right)^{p+2} \left(m-p\right) \left(n-p\right) \dfrac{m+n-2p+1+\nu}{m+n-2p+1} \dfrac{m+n-2p-1+\nu}{m+n-2p-1}
y_{\alpha\left(m-1-p\right)} y_{\beta\left(n-1-p\right)} \right)
\end{multline}
Shifting the indices in the sums we can turn it in the form of Eq.~(\ref{Ferm}), where $B\left(m+1,n+1,p,\nu\right)$ in terms of bosonic structure constants is

\begin{multline}
B\left(m+1,n+1,p,\nu\right)=A\left(m,n,p,-\nu\right)+i \left(m+n-2p+3+\nu\right)A\left(m,n,p-1,-\nu\right)+\\
+i^2 \left(m-p+2\right)\left(n-p+2\right) \dfrac{m+n-2p+5+\nu}{m+n-2p+5} \dfrac{m+n-2p+3+\nu}{m+n-2p+3} A\left(m,n,p-2,\nu\right) .
\end{multline}

\section{$Even \times Odd$ and $Odd \times Even$}

Just as in the previous section one may derive structure constants for \textit{Even $\times$ Odd} and \textit{Odd $\times$ Even} cases. We skip algebra and give only the final answers:

\textit{Even $\times$ Odd}

\begin{equation}
C\left(m,n+1,p,\nu\right)=A\left(m,n,p,-\nu\right)+i\left(m-p+1\right)\dfrac{m+n-2p+3+\nu}{m+n-2p+3} A\left(m,n,p-1,-\nu\right),
\end{equation}

\textit{Odd $\times$ Even}

\begin{equation}
D\left(m+1,n,p,\nu\right)=A\left(m,n,p,\nu\right)+i\left(n-p+1\right)\dfrac{m+n-2p+3-\nu}{m+n-2p+3} A\left(m,n,p-1,\nu\right).
\end{equation}

\section{Conclusion}
All structure constants for $Aq\left(2,\nu\right)$ in covariant basis are found. Associativity of this product is proved. The fact that structure constants are Saalschutzian hypergeometric functions is crucial. Saalschutzian transform is the key ingredient that allows to show that associativity condition holds.

\section{Acknowledgements}
The author is grateful to Mikhail Vasiliev for useful comments on the manuscript and acknowledges financial support from Dynasty Foundation. This research was supported by RFBR Grant No 14-02-01172.

\section*{Appendix. Associativity check}
In \cite{Pope} associative product was given in the form

\begin{equation}
\label{LS}
V_m^s \ast V_n^t=\dfrac{1}{2} \sum_{u=1}^{s+t-1} g_u^{st} \left(m,n,\lambda\right) V^{s+t-u}_{m+n},
\end{equation}
with
\begin{multline}
\displaystyle
g_u^{st} \left(m,n,\lambda\right)=\dfrac{\left(\frac{1}{4}\right)^{u-2}}{2\left(u-1\right)!} {}_4F_3\left[\begin{matrix} \frac{1}{2}+\lambda & \frac{1}{2}-\lambda &\frac{2-u}{2}& \frac{1-u}{2}\\
\frac{3-2s}{2}& \frac{3-2t}{2} &\frac{1}{2}+s+t-u \end{matrix};1\right] \times \\
\sum_{k=0}^{u-1} \left(-1\right)^k \left(\begin{matrix}u-1 \\ k \end{matrix}\right) \left(s-1-m\right)_{u-1-k}\left(s-1-m\right)_k\left(t-1+n\right)_k \left(t-1-n\right)_{u-1-k},
\end{multline}
where $\left(a\right)_n$ is the descending Pochhammer symbol. In \cite{Pope} associativity of Lone-Star product was only conjectured, in this paper we explicitly check it. To get the covariant expression from (\ref{LS})  we have to multiply highest and lowest state vectors, in this case all the contractions with $\epsilon_{\alpha\beta}$ are non-zero. After shifting indices and plugging $\nu=1-2\lambda$ we get the expression (\ref{Aprod}). We need to plug it then into associativity  Eq. (\ref{eq0}, \ref{eq1},\ref{main}, \ref{eq2}, \ref{eq3}). Because factorial term provides all necessary zeros we may use only Eq.~(\ref{main}). For shortness we denote
\begin{equation}
{}_4 F_3\left[\begin{matrix} 1-\frac{\nu}{2}& \frac{\nu}{2} &\frac{-p}{2}& \frac{1-p}{2}\\
\frac{1-m}{2}& \frac{1-n}{2} &\frac{m+n-2p+3}{2}\end{matrix};1\right]=F\left(m , n , p , \nu\right)
\end{equation}
Identity it should satisfy
\begin{multline}
\dfrac{n!}{\left(n-p\right)! p!} F\left(m,n,p,\nu\right)=\dfrac{\left(n-2\right)!}{\left(n-p-2\right)! p!} F\left(m,n-2,p,\nu\right)+\\
+2 \dfrac{\left(n-2\right)!}{\left(n-p-1\right)! \left(p-1\right)!} F\left(m,n-2,p-1,\nu\right)+\\
+\dfrac{\left(n-2\right)!}{\left(n-p\right)! \left(p-2\right)!} \dfrac{m+n-2p+3-\nu}{m+n-2p+3}\ \dfrac{m+n-2p+1+\nu}{m+n-2p+1} F\left(m,n-2,p-2,\nu\right).
\end{multline}
It is better to treat Hypergeometries from r.h.s. separately.

\subsection*{First term $F\left(m,n-2,p,\nu\right)$ }

According to definition it can be written as a series

\begin{multline}
\displaystyle
F\left(m,n-2,p,\nu\right)={}_4 F_3\left[\begin{matrix} 1-\frac{\nu}{2}& \frac{\nu}{2} &\frac{-p}{2}& \frac{1-p}{2}\\
\frac{1-m}{2}& \frac{1-n}{2}+1 &\frac{m+n-2p+3}{2}-1\end{matrix};1\right]=\\
=\sum_{q=0}^{\infty}\dfrac{\left(1-\frac{\nu}{2}\right)_q \left(\frac{\nu}{2}\right)_q \left(\frac{1-p}{2}\right)_q\left(-\frac{p}{2}\right)_q}{\left(\frac{1-m}{2}\right)_q \left(\frac{1-n}{2}+1\right)_q \left(\frac{m+n-2p+3}{2}-1\right)_q q!}.
\end{multline}
Performing transformation of Pochhammer symbols we can obtain in the form of $F\left(m,n,p,\nu\right)$ series. Using the definition of Pochhammer symbols
\begin{equation}
\left(a\right)_q=\dfrac{\Gamma\left(a+q\right)}{\Gamma\left(a\right)}
\end{equation}
we proceed with a new expression for  $F\left(m,n-2,p,\nu\right)$

\begin{multline}
F\left(m,n-2,p,\nu\right)=\dfrac{m+2n-2p}{m+n-2p+1}{}_4 F_3\left[\begin{matrix} 1-\frac{\nu}{2}& \frac{\nu}{2} &\frac{-p}{2}& \frac{1-p}{2}\\
\frac{1-m}{2}& \frac{1-n}{2}+1 &\frac{m+n-2p+3}{2}\end{matrix};1\right]-\\
-\dfrac{n-1}{m+n-2p+1}
{}_4 F_3\left[\begin{matrix} 1-\frac{\nu}{2}& \frac{\nu}{2} &\frac{-p}{2}& \frac{1-p}{2}\\
\frac{1-m}{2}& \frac{1-n}{2} &\frac{m+n-2p+3}{2}\end{matrix};1\right].
\end{multline}
Second term is indeed hypergeometry from l.h.s. of associativity Eq.~(\ref{main}).

\subsection*{Third term $F\left(m,n-2,p-2,\nu\right)$}

Here the fact that our ${}_4 F_3 \left(1\right)$ are Saalschutzian is used and the transformation is performed. More information about terminating Saalschutzian series can be found in \cite{Slater}. Let $p=2N$, odd case can be treated in the same way.

\begin{multline}
F\left(m,n-2,p-2,\nu\right)={}_4 F_3\left[\begin{matrix} 1-\frac{\nu}{2}& \frac{\nu}{2} &1-\frac{p}{2}& 1+\frac{1-p}{2}\\
\frac{1-m}{2}& \frac{1-n}{2}+1 &\frac{m+n-2p+3}{2}+1\end{matrix};1\right]=\\
=\dfrac{\left(\frac{p-m}{2}-1\right)_{N-1}}{\left(\frac{1-m}{2}\right)_{N-1}} \dfrac{\left(\frac{p-n}{2}\right)_{N-1}}{\left(\frac{1-n}{2}+1\right)_{N-1}} {}_4 F_3\left[\begin{matrix} 1+\frac{m+n-2p+1+\nu}{2}& 1+\frac{m+n-2p+3-\nu}{2} &1-\frac{p}{2}& 1+\frac{1-p}{2}\\
\frac{m}{2}-p+3& \frac{n}{2}-p+2 &\frac{m+n-2p+3}{2}+1\end{matrix};1\right].
\end{multline}
We need to extract the series that will coincide with hypergeometry form l.h.s. of associativity equation. After simple algebra with shifting the index in the sum and transformation of Pochhammer symbols(like with the first term) we proceed with new expression for the third term

\begin{multline}
F\left(m,n-2,p-2,\nu\right)=\dfrac{\left(\frac{p-m}{2}-1\right)_{N-1}}{\left(\frac{1-m}{2}\right)_{N-1}} \dfrac{\left(\frac{p-n}{2}\right)_{N-1}}{\left(\frac{1-n}{2}+1\right)_{N-1}} \dfrac{\left(\frac{m+n-2p+3}{2}\right)\left(\frac{m}{2}-p+2\right)\left(\frac{n}{2}-p+1\right)}{\left(\frac{m+n-2p+3-\nu}{2}\right)\left(\frac{m+n-2p+1+\nu}{2}\right)\left(\frac{1-p}{2}\right)\left(\frac{-p}{2}\right)} \times \\ \left(\frac{m}{2}-p+1\right)
  \left( {}_4 F_3\left[\begin{matrix} \frac{m+n-2p+1+\nu}{2}& \frac{m+n-2p+3-\nu}{2} &-\frac{p}{2}& \frac{1-p}{2}\\
\frac{m}{2}-p+1& \frac{n}{2}-p+1 &\frac{m+n-2p+3}{2}\end{matrix};1\right]- \right. \\ \left. -{}_4 F_3\left[\begin{matrix} \frac{m+n-2p+1+\nu}{2}& \frac{m+n-2p+3-\nu}{2} &-\frac{p}{2}& \frac{1-p}{2}\\
\frac{m}{2}-p+2& \frac{n}{2}-p+1 &\frac{m+n-2p+3}{2}\end{matrix};1\right]  \right).
\end{multline}

\subsection*{Second term $F\left(m,n-2,p-1,\nu\right)$}

It is more convenient to treat the second term in the end. After transformation of the 1st and 3rd we get Hypergeometries from l.h.s. of Eq.~(\ref{main}) plus some extra terms. We split 2nd term into two
\begin{multline}
F\left(m,n-2,p-1,\nu\right)=A\; F\left(m,n-2,p-1,\nu\right) + B\; F\left(m,n-2,p-1,\nu\right) \\
A+B=1.
\end{multline}
\textit{A}-term will transformed as 1st Hypergeometry, \textit{B} term as 3rd. Constants $A$ and $B$ will be chosen to cancel extra terms from 1st and 3rd term. Transformed \textit{A}-term has the form

\begin{multline}
F\left(m,n-2,p-1,\nu\right)=\frac{n-1}{p} {}_4 F_3\left[\begin{matrix} 1-\frac{\nu}{2}& \frac{\nu}{2} &\frac{-p}{2}& \frac{1-p}{2}\\
\frac{1-m}{2}& \frac{1-n}{2} &\frac{m+n-2p+3}{2}\end{matrix};1\right]-\\
-\frac{\left(n-p-1\right)}{p} {}_4 F_3\left[\begin{matrix} 1-\frac{\nu}{2}& \frac{\nu}{2} &\frac{-p}{2}& \frac{1-p}{2}\\
\frac{1-m}{2}& \frac{1-n}{2}+1 &\frac{m+n-2p+3}{2}\end{matrix};1\right].
\end{multline}
Transformed \text{B}-term

\begin{multline}
F\left(m,n-2,p-1,\nu\right)=
\dfrac{\left(\frac{p-m}{2}\right)_N}{\left(\frac{1-m}{2}\right)_N} \dfrac{\left(\frac{p-n}{2}\right)_N}{\left(\frac{1-n}{2}\right)_N} \dfrac{n-1}{\left(n-p\right) p} \times \\
\times \left( \dfrac{\left(\frac{m-p+1}{2}\right)\left(m-p+2\right)}{\frac{m}{2}-p+1} {}_4 F_3\left[\begin{matrix} \frac{m+n-2p+1+\nu}{2}& \frac{m+n-2p+3-\nu}{2} &-\frac{p}{2}& \frac{1-p}{2}\\
\frac{m}{2}-p+2& \frac{n}{2}-p+1 &\frac{m+n-2p+3}{2}\end{matrix};1\right]-\right.\\
\left. - \left(m-p+1\right){}_4 F_3\left[\begin{matrix} \frac{m+n-2p+1+\nu}{2}& \frac{m+n-2p+3-\nu}{2} &-\frac{p}{2}& \frac{1-p}{2}\\
\frac{m}{2}-p+1& \frac{n}{2}-p+1 &\frac{m+n-2p+3}{2}\end{matrix};1\right] \right).
\end{multline}
Values of $A$ and $B$, when extra terms vanish

\begin{equation}
A=\dfrac{m+2n-2p}{m+n-2p+1}, \;\;\;
B=1 - \dfrac{m+2n-2p}{m+n-2p+1}.
\end{equation}
Plugging all three terms in associativity Eq.~(\ref{main}) we see that it is not violated.

\end{document}